\documentclass[prl, aps, reprint, superscriptaddress, preprintnumbers, amsfonts, amssymb, amsmath%
, noeprint%
]{revtex4-1}
\usepackage{mathrsfs}
\usepackage{graphicx}
\usepackage{hyperref}
\usepackage{slashed}
\hypersetup{
colorlinks, citecolor=[rgb]{0,0,0.8}, linkcolor=[rgb]{0,0,0.8}, urlcolor=[rgb]{0.7,0,0.5}}

\newcommand{\wb}{\overline}

\newcommand{\eg}{\textit{e.g.}}

\newcommand{\ie}{\textit{i.e.}}

\newcommand{\nn}{\nonumber}

\newcommand{\be}{\begin{equation}} \newcommand{\ee}{\end{equation}}
\newcommand{\bea}{\begin{equation} \begin{aligned}} \newcommand{\eea}{\end{aligned} \end{equation}}

\newcommand{\cC}{\mathcal{C}}

\newcommand{\cL}{\mathcal{L}}

\newcommand{\cO}{\mathcal{O}}

\DeclareMathOperator{\Tr}{Tr}

\begin{document}

\title{Conformality Loss, Walking, and 4D Complex \\[.1em] Conformal Field Theories at Weak Coupling}

\date{\today}

\author{Francesco Benini}
\email{fbenini@sissa.it}

\affiliation{SISSA, via Bonomea 265, 34136 Trieste, Italy and \\ INFN, Sezione di Trieste, via Valerio 2, 34127 Trieste, Italy}
\affiliation{ICTP, Strada Costiera 11, 34151 Trieste, Italy}

\author{Cristoforo Iossa}
\email{ciossa@sissa.it}

\affiliation{SISSA, via Bonomea 265, 34136 Trieste, Italy and \\ INFN, Sezione di Trieste, via Valerio 2, 34127 Trieste, Italy}
\affiliation{University of Trento, Via Sommarive 14, 38123 Trento, Italy}

\author{Marco Serone}
\email{serone@sissa.it}

\affiliation{SISSA, via Bonomea 265, 34136 Trieste, Italy and \\ INFN, Sezione di Trieste, via Valerio 2, 34127 Trieste, Italy}
\affiliation{ICTP, Strada Costiera 11, 34151 Trieste, Italy}

\begin{abstract}
Four-dimensional gauge theories with matter can have regions in parameter space, often dubbed conformal windows, where they flow in the infrared to non-trivial conformal field theories. It has been conjectured that conformality can be lost because of merging of two nearby fixed points that move into the complex plane, and that a walking dynamics governed by scaling dimensions of operators defined at such complex fixed points can occur. We find controlled, parametrically weakly coupled, and ultraviolet-complete 4D gauge theories that explicitly realize this scenario. We show how the walking dynamics is controlled by the coupling of a double-trace operator that crosses marginality. The walking regime ends when the renormalization group flow of this coupling leads to a (weak) first-order phase transition with Coleman-Weinberg symmetry breaking. A light dilaton-like scalar particle appears in the spectrum, but it is not parametrically lighter than the other excitations.
\end{abstract}


\preprint{SISSA 21/2019/FISI}

\maketitle

Understanding the phases of gauge theories is one of the most interesting problems in high energy and condensed matter physics. 
Ultraviolet(UV)-free gauge theories with matter exhibit so-called conformal windows, namely regions in parameter space where they flow in the infrared (IR) to interacting conformal field theories (CFTs).  A relevant example is given by four-dimensional (4D)
$SU(N_c)$ gauge theories with $N_f$ fermions in the fundamental representation of the gauge group. At fixed $N_c$, the conformal window spans
an interval \mbox{$N_f^{-} \leq N_f \leq N_f^{+}$}. For values of $N_f$ outside this range, IR conformality is lost and it is interesting to understand the mechanism of how this happens.
The upper edge of the conformal window, $N_f^{+} = \frac{11}2 N_c$, is more accessible because the theory can be weakly coupled there. 
When $N_f \to \raisebox{0pt}[0pt][0pt]{$N_f^+$}$ the non-trivial fixed point collides with the Gaussian one, and the latter changes its stability for $N_f > N_f^+$ (\ie, the theory is no longer UV free).
The lower edge is instead strongly coupled and difficult to analyze.
It has been proposed that conformality is lost because two nearby fixed points merge and disappear into the complex plane \cite{Kaplan:2009kr} (see also \cite{Gies:2005as}). 
Just below the lower edge of the conformal window, a related interesting phenomenon is expected to occur \cite{Kaplan:2009kr}: walking dynamics, namely 
a long range of energy scales where the renormalization group (RG) flow slows down and the theory is approximately conformal invariant, after which chiral symmetry breaking and confinement take place
\cite{Holdom:1981rm, *Yamawaki:1985zg, *Appelquist:1986an, *Appelquist:1996dq}%
\footnote{Walking dynamics has been mostly discussed within technicolor models (see \cite{Hill:2002ap} for a review and further references), but it can play an important role also in its modern reincarnation in terms of UV-complete composite Higgs models, as in \cite{Ferretti:2013kya}.}.
The walking behavior would occur because the RG flow passes ``between'' and close to the two \emph{complex} fixed points and is governed by scaling dimensions of operators in the (non-unitary) CFTs living at those fixed points \cite{Gorbenko:2018ncu, Gorbenko:2018dtm}.
Similar dynamics has been conjectured to take place also in three-dimensional gauge theories \cite{Appelquist:1988sr, *Appelquist:1989tc, *Nahum:2015jya}.

Conformality loss and walking in 4D strongly coupled gauge theories are still conjectural.
It would be interesting to confidently establish their existence, at least in weakly coupled models.
In order to claim success, several requirements should be met. In particular, these theories should be
(1) unitary; (2) UV complete; (3) stable
\footnote{Stability means that the scalar potential is bounded below, to give a consistent and unitary quantum theory.}.
In addition, they should have a parameter that makes the full RG flow arbitrarily weakly coupled (to make the perturbative expansion reliable), and another one that allows us to tune from a regime with an IR CFT to a regime with complex fixed points and walking.

In this Letter we show how the above requirements can be fulfilled in a specific (and simple) class of models.
Being weakly coupled, those models will also be calculable.

When two fixed points collide, the operator $\cO_f$ governing the flow from one to the other becomes marginal. Let us denote by $f$ the coupling associated to $\cO_f$. Up to rescaling and shift, for small $f$ its $\beta$-function reads $\beta_f = - y + f^2 + O(f^3)$,
where $y$ is a small parameter, possibly dependent on other couplings and parameters in the theory. For $y>0$ there are two fixed points, at $y=0$ they merge, and for $y<0$ they become complex.
In large $N$ theories, $\cO_f$ should be a double-trace operator $\cO^2$ \cite{Gubser:2002vv} (see also Appendix D of \cite{Gorbenko:2018ncu}). 
At leading order in the large $N$ limit, but to all orders in $f$ and other single-trace deformation couplings $\lambda$ (such as the 't~Hooft gauge coupling), the $\beta$-functions read \cite{Witten:2001ua, Dymarsky:2005uh, *Pomoni:2008de}\footnote{The possibility of a walking behavior was discussed  in \cite{Dymarsky:2005uh} in the context of orbifold models of ${\cal N}=4$ supersymmetric Yang-Mills theory, where $b(\lambda)$ in (\ref{betaflambda}) vanishes identically. However, no real fixed points were found, and such theories are generally not UV complete.}
\be
\label{betaflambda}
\beta_f  =  c_1(\lambda) \, f^2 + c_2 (\lambda) \, f + c_3 (\lambda)  \;,\qquad
\beta_\lambda   =   b(\lambda) \;.
\ee
This applies also in the presence of several single-trace deformations and in more general large $N$ limits, such as the Veneziano limit. 
If the theory flows to an IR fixed point, (\ref{betaflambda}) implies that the double-trace operator $\cO^2$ will not mix with 
the single-trace operators with couplings $\lambda_i$, and will have scaling dimension $\Delta_{\cO^2} = 4 + \gamma_{\cO^2}$ with $\gamma_{\cO^2} = \partial_f \beta_f |^\star =  2f^\star c_1(\lambda^\star) + c_2(\lambda^\star) = 2\gamma_{\cO} + O(1/N)$.
Here the star superscript indicates the value at the fixed point. At weak coupling $\Delta_{{\cal O}} \approx 2$ and hence we are forced to introduce scalar fields.
By varying some parameter in the theory, we can expect to reach a point where $\gamma_{\cO^2}$ vanishes, possibly signaling fixed-point merging.
All these considerations lead us to consider large $N$ gauge theories that include scalar fields and a double-trace deformation.

A simple model of this kind is given by an $SU(N_c)$ gauge theory with $N_f$ Dirac fermions and $N_s$ scalars in the fundamental representation
\footnote{A numerical study of this theory at finite $N$, including its RG flow, regions of conformality and UV freedom, already appeared in \cite{Hansen:2017pwe}. However, an analytic large $N$ study, description of fixed-point merging and walking behaviour was not provided there.}.
The Lagrangian is
\begin{multline}
\cL = - \tfrac14 F_{\mu\nu}^A F^{\mu\nu}_A + \Tr \bar\psi i \slashed{D} \psi + \Tr D_\mu \phi^\dag D^\mu \phi \\
{} - \tilde h \Tr \phi^\dag \phi \phi^\dag \phi - \tilde f \Tr \phi^\dag \phi \Tr \phi^\dag \phi
\end{multline}
(before gauge fixing).
The gauge coupling $g$ is in the covariant derivative.
As we will see, the theory is UV complete and free and hence technically natural: there exist renormalization schemes (\eg, minimal subtraction in dimensional regularization, $\overline{\text{MS}}$) in which the scalar masses remain zero to all orders in perturbation theory. Fermion masses are protected by the $SU(N_f)^2$ chiral symmetry.

We are interested in finding weakly coupled Caswell-Banks-Zaks (CBZ) fixed points \cite{Caswell:1974gg, *Banks:1981nn}. This requires us to take a large $N$ limit. We introduce 't~Hooft couplings
\be
\lambda = \frac{N_c \, g^2}{16\pi^2} \;,\qquad h = \frac{N_c \, \tilde h}{16\pi^2} \;,\qquad f = \frac{N_c N_s \, \tilde f}{16\pi^2}
\ee
and take the Veneziano limit $N_c, N_f, N_s \to \infty$ with
\be
x_s = {N_s}/{N_c} \;,\qquad\qquad x_f = {N_f}/{N_c}
\ee
and $\lambda, h, f$ fixed. Notice the different scaling limit of $h$ and $f$ with $N_c$, reflecting the different single- and double-trace nature of the corresponding operators.
The perturbative $\beta$-functions, at two-loop order in the gauge coupling and one-loop order in the other couplings, are \cite{Machacek:1983tz, *Machacek:1984zw}
\begin{subequations}
\label{beta functions}
\begin{align}
\beta_\lambda &= - \frac{22 - x_s - 4 x_f}3 \, \lambda^2 + b_1 \lambda^3 \\
\beta_h &= 4 (1+x_s) h^2 + \frac{24}{N_cN_s} fh \nn \\
&\quad - \Bigl(6 - \frac6{N_c^2} \Bigr) \lambda h + \Bigl( \frac34 - \frac3{N_c^2} \Bigr) \lambda^2 \\
\beta_f &= \Bigl( 4 + \frac{16}{N_cN_s} \Bigr) f^2 + 8 (1+x_s) fh + 12 x_s h^2 \nn \\
&\quad - \Bigl( 6 - \frac6{N_c^2} \Bigr) \lambda f + \frac{3x_s}4 \Bigl( 1 + \frac2{N_c^2} \Bigr) \lambda^2
\end{align}
\end{subequations}
where the gauge coefficient $b_1$ is
\be
b_1 = \frac23 \bigl( 4x_s + 13x_f - 34 \bigr) - \frac{2(x_s + x_f)}{N_c^2} \;.
\ee
These $\beta$-functions are in agreement with the general form  (\ref{betaflambda}).
The two-loop coefficients are scheme dependent when multiple couplings are present; the result reported here is in $\overline{\text{MS}}$. We study the theory at leading order in $1/N_c$ and in a loopwise expansion, meaning that parametrically $\lambda \sim h \sim f$. In the CBZ limit, the coefficient multiplying the $\lambda^2$ term in $\beta_\lambda$ is made artificially small, justifying the use of two-loop order in the gauge but not the other couplings.
In order to obtain a weakly coupled fixed point, we set
\be
22 - x_s - 4 x_f = 75 \epsilon
\ee
and take $0 < \epsilon \ll 1$. The numerical factor in front of $\epsilon$ has been chosen for later convenience. Note that $\epsilon$ can be made as small as of order $1/N_c$. We will use $x_s$ to parametrize a family of CBZ fixed points, working at leading order in $1/N_c$.
The non-trivial zero of $\beta_\lambda$ is at
\be
\label{lambda star}
\lambda^\star = \frac{\epsilon}{1 + x_s/50 - 13\epsilon/2} \;.
\ee
Taking $\epsilon \lesssim 0.1$, $\lambda^\star$ is positive, of order $\epsilon$ and small.
Plugging that value into $\beta_h$, the latter has two zeros at
\be
\label{h star}
h^\star_\pm = \lambda^\star \, \frac{ 3 \pm \sqrt{6-3x_s} }{ 4(1+x_s) } \;.
\ee
Both values are real positive for $0\leq x_s \leq 2$. Finally, plugging $\lambda^\star, h^\star_\pm$ into $\beta_f$, the latter has four zeros at
\bea
\label{f star}
f^\star_{\pm+} &\equiv f^\star_\pm(h^\star_+) = \lambda^\star \, ( -B \pm A_+) \\
f^\star_{\pm-} &\equiv f^\star_\pm(h^\star_-) = \lambda^\star \, ( +B \pm A_-)
\eea
with $B= \sqrt{6 - 3x_s}/4$ and 
\be
\label{A and B}
A_\pm = \frac{ 3 \sqrt{ 2 - \bigl( 13 \pm 6 \sqrt{ 6-3x_s} \bigr) x_s + x_s^2 -2x_s^3 } }{ 4 \sqrt3 \, (1+x_s) } \;.
\ee
For $x_s<2$, $B$ is real positive. On the other hand, $A_\pm$ are real (and non-negative) in a smaller range: $A_+$ for
\be
x_s \leq \wb x_s =0.07309 \;,
\ee
while $A_-$ for $x_s \leq 0.8403$ (the numerical values are approximate). Outside those ranges, $A_\pm$ are complex.

\begin{figure}[t]
\centering
\includegraphics[width=0.7\columnwidth]{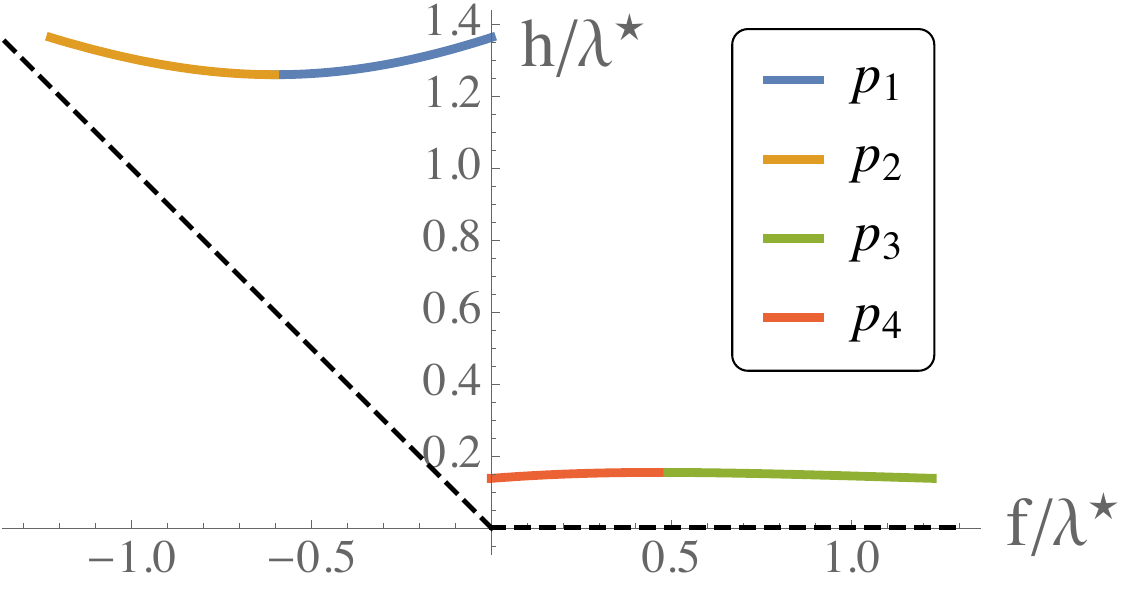}
\caption{Positions of the IR fixed points $p_j$ as we vary $x_s$. The couplings are apart at $x_s=0$ and progressively merge ($p_{1,2}$ at $x_s = \wb x_s$ while $p_{3,4}$ at $x_s \simeq 0.8$) as we increase $x_s$. The region above the dashed lines corresponds to positive potentials.
\label{fig: couplings}}
\end{figure}

We thus find four interacting fixed points $p_j$ (plotted in Fig.~\ref{fig: couplings} as $x_s$ is varied), with couplings
\bea
p_1 &= \{ \lambda^\star, h^\star_+, f^\star_{++}\} \;,\qquad& p_3 &= \{ \lambda^\star, h^\star_-, f^\star_{+-} \} \\
p_2 &= \{ \lambda^\star, h^\star_+, f^\star_{-+} \} \;,\qquad& p_4 &= \{ \lambda^\star, h^\star_-, f^\star_{--} \} \;.
\eea
The stability properties under RG flow of the four fixed points are determined by the eigenvalues of the matrix $\partial \beta_j/ \partial c_k$ where $\beta= (\beta_\lambda, \beta_h, \beta_f)$ and $c = (\lambda, h, f)$. The three eigenvalues are
\bea
\zeta_1 &= -50\epsilon\lambda + 75 \bigl(1+x_s/50 - 13\epsilon/2 \bigr) \lambda^2 \\
\zeta_2 &= 8(1+x_s)h -6\lambda \;,\qquad\quad \zeta_3 = \zeta_2 + 8f \;.
\eea
We evaluate these eigenvalues at the fixed points. The first eigenvalue is the same at the four fixed points, namely $\zeta_1^\star = 25 \epsilon \lambda^\star$.
Since it is positive, this direction is always IR stable. The other two eigenvalues take the following simple form.
\bea
\text{At } (h^\star_+, f^\star_{\pm+})\!: \quad &\zeta_2^\star = +8\lambda^\star B \;,\quad \zeta_3^\star = \pm 8\lambda^\star A_+ \\
\text{At } (h^\star_-, f^\star_{\pm-})\!: \quad &\zeta_2^\star = -8\lambda^\star B \;,\quad \zeta_3^\star = \pm 8 \lambda^\star A_- \;.
\eea
We conclude that, in the domain where such fixed points are real, $p_1$ has two IR-stable directions, $p_{2,3}$ have a stable and an unstable one, while $p_4$ has two unstable directions (besides $\zeta_1^\star$).
Moreover, $p_1$ and $p_2$ merge at $x_s = \wb x_s$ where \mbox{$\lambda_3^\star=0$}, and for larger values of $x_s$ they move into the complex plane. Similarly, $p_3$ and $p_4$ merge at $x_s \simeq 0.8$ where $\lambda_3^\star=0$ and
then move to the complex plane.
The linear combinations of operators that diagonalize dilations correspond to eigenvectors of the matrix $\partial \beta_j / \partial c_k$. As expected, the double-trace operator $\cO^2 = \bigl( \Tr \phi^\dag \phi \bigr){}^2$ does not mix at leading order in $1/N_c$ with the two single-trace operators $\Tr F_{\mu\nu} F^{\mu\nu}$ and $\Tr \phi^\dag\phi \phi^\dag \phi$. 
The dimension of $\cO^2$ is $\Delta_{\cO^2} = 4 + \zeta_3^\star$. At the values of $x_s$ where fixed points merge and $\zeta_3^\star$ vanishes, the double-trace operator becomes marginal, again as expected. 

The scalar potential $V =  \tilde h \Tr \phi^\dag\phi \phi^\dag \phi + \tilde f \, \bigl( \Tr \phi^\dag\phi \bigr){}^2$ is positive definite at the fixed points. Indeed, one can prove that this is the case if and only if $\tilde f + \tilde h>0$ and $M \tilde f + \tilde h>0$, where $M = \min(N_c,N_s)$. In the large $N$ limit, these conditions become
\be
\label{eq:stabilitycond}
h>0 \qquad\text{and}\qquad \min\bigl(1, x_s^{-1}\bigr)\, f + h > 0 \;.
\ee
All fixed points $p_{1,2,3,4}$ satisfy the positivity conditions, for all values of $x_s$ for which they are real (see Fig.~\ref{fig: couplings}).

We inquire whether the model defines a UV complete quantum field theory, namely, whether the theory is UV free. In order to study the RG flow around the origin in the space $(\lambda, h, f)$, it suffices to use one-loop $\beta$-functions. The analysis is similar to \cite{Hansen:2017pwe}. First, we identify special flow lines that we call ``radial'':
\be
\lambda = \gamma_\lambda \, \rho(\mu) \;,\quad h = \gamma_h \, \rho(\mu) \;,\quad f = \gamma_f \, \rho(\mu)
\ee
where $\rho(\mu)$ is a positive function of the renormalization scale $\mu$, while $\gamma_j$ are constants (defined up to positive rescalings). 
At leading order in $\epsilon$, four radial flows are given by $\gamma_h/ \gamma_\lambda \,\simeq\, h^\star_\pm/ \lambda^\star$, $\gamma_f / \gamma_\lambda \,\simeq\, f^\star_{\pm\pm}/\lambda^\star$, in terms of the IR values in (\ref{lambda star})--(\ref{f star}). 
Extra radial flows lie on the plane $\lambda=0$ (\ie, have $\gamma_\lambda=0$) with $\gamma_h=0$ or $\gamma_f/\gamma_h = - \bigl( 1+x_s \pm \sqrt{1-10x_s+x_s^2} \bigr)/2$.

\begin{figure}[t]
\centering
\includegraphics[height=.49\columnwidth]{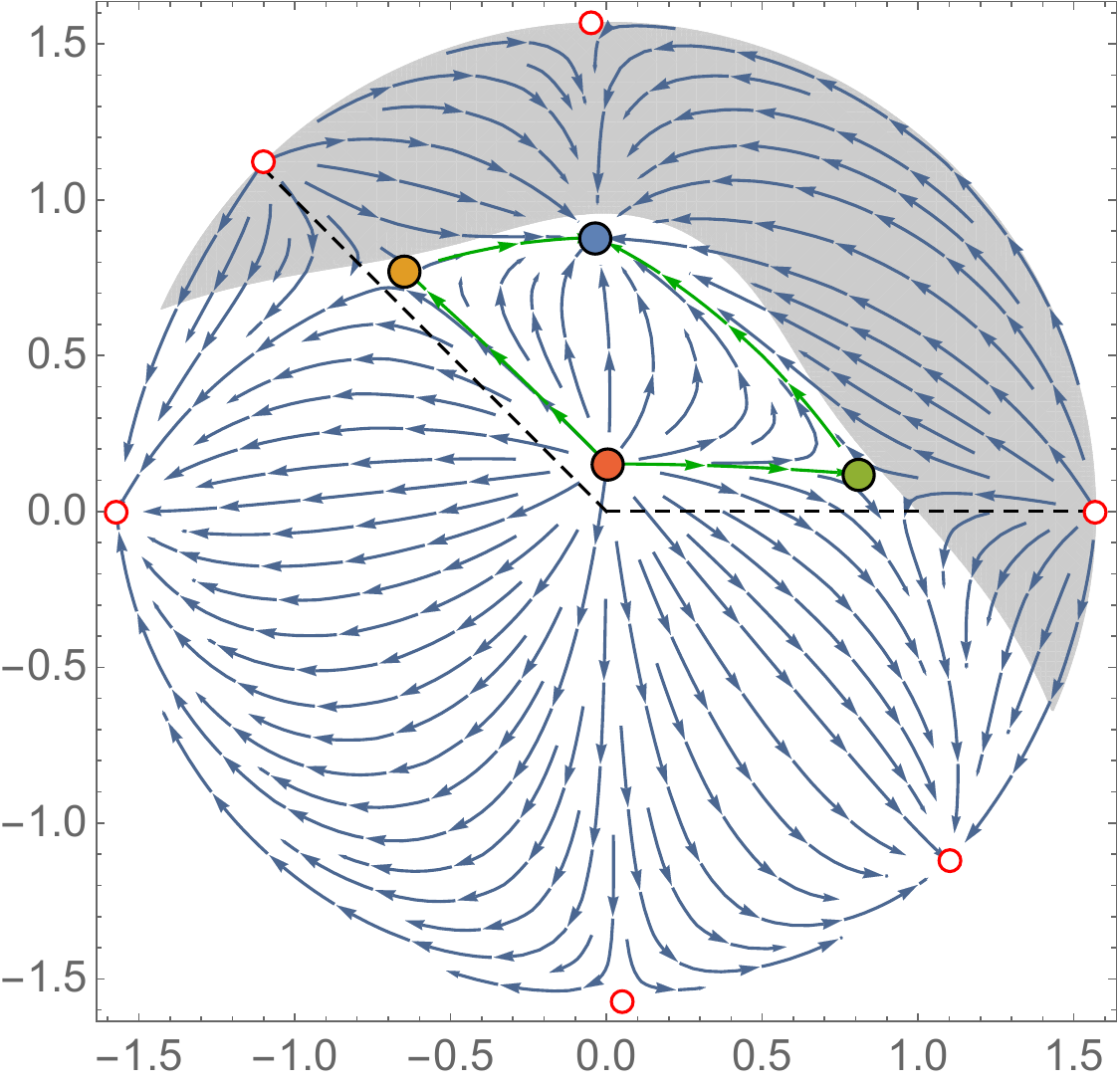}
\includegraphics[height=.49\columnwidth]{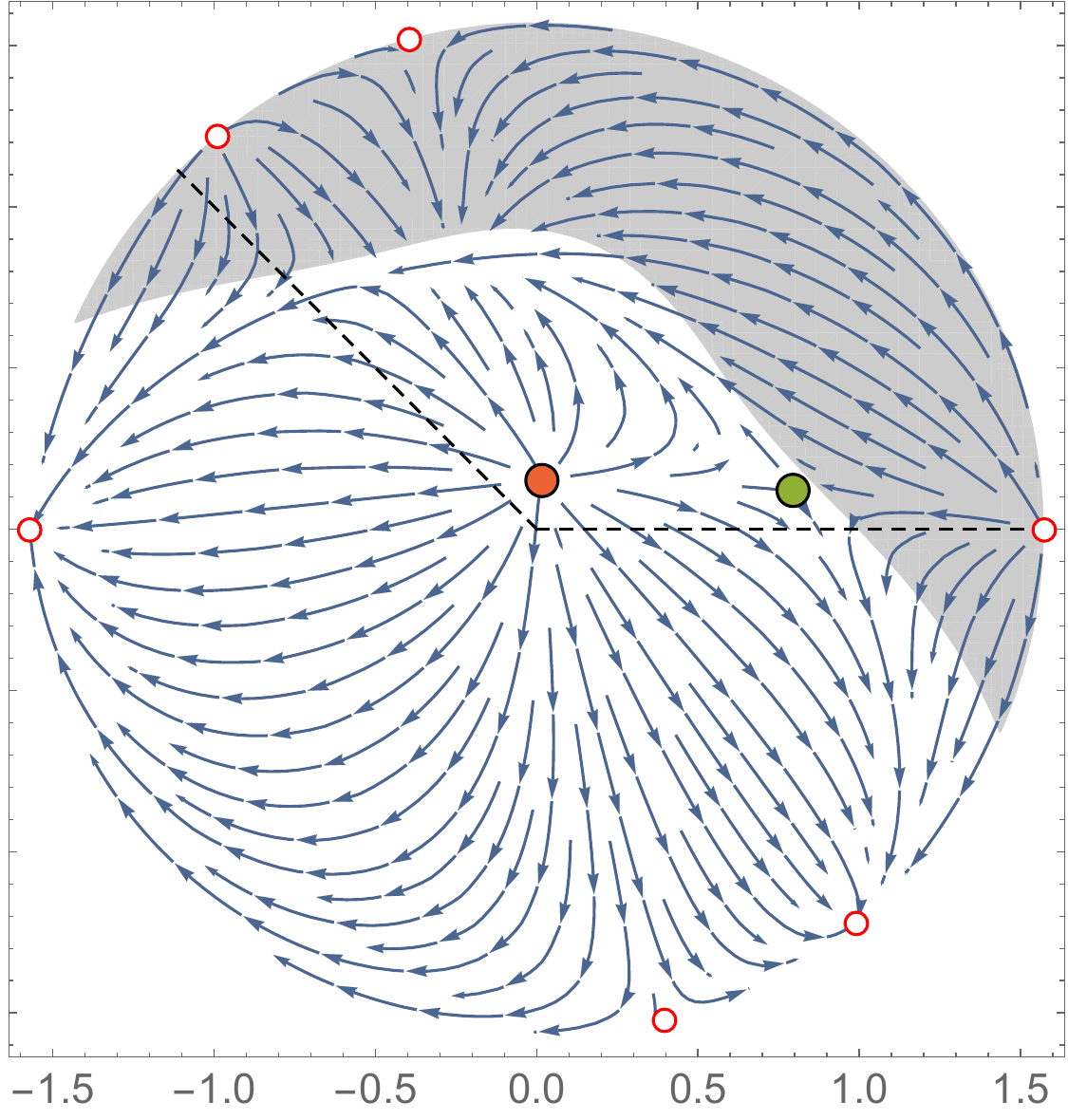}
\caption{UV angular flows on the hemisphere $0\leq \theta\leq \frac\pi2$, where $(\theta,\varphi)$ are used as two-dimensional polar coordinates. Arrows point towards the IR. In both figures $\epsilon = 0.02$. Left: $x_s=0.01$. Right: $x_s = \wb x_s$.
\label{fig: angular flows}}
\end{figure}

Then, we introduce spherical coordinates $(r,\theta,\varphi)$ in the space of couplings, setting
\be
\lambda = r \cos\theta \;,\quad h = r \sin\theta \sin\varphi \;,\quad f = r \sin\theta \cos\varphi
\ee
with $\theta \in [0, \frac\pi2]$ ($\lambda \geq 0$) and $\varphi \in [0,2\pi)$. Correspondingly, we define radial and angular $\beta$-functions $\beta_r$ and $\beta_\theta, \beta_\varphi$. Since the one-loop $\beta$-functions in (\ref{beta functions}) are homogeneous in the couplings, the angular functions $\beta_\theta, \beta_\varphi$ are proportional to $r$ and can be studied separately from $\beta_r$. Their fixed points are precisely the radial flows discussed above.

We plot the UV angular flows, for different values of $x_s$, in Fig.~\ref{fig: angular flows}. The angular flow takes place on a hemisphere at fixed $r$, and we use $(\theta, \varphi)$ as two-dimensional polar coordinates. The gray shaded region is where $\beta_r>0$. The region above the dashed line is where the potential $V$ is positive definite. Radial flows are represented by dots: white dots are radial flows in the $\lambda=0$ plane; blue, yellow, green and red dots are the other ones
(that, for $\epsilon\to0$, correspond to the IR fixed points $p_j$ in Fig.~\ref{fig: couplings}).
The red dot is the UV-attractive radial flow that makes the theory UV free. The theory is UV free for small values of $x_s$ as long as $\epsilon \lesssim 0.1$, and is UV free up to the IR merging point $x_s = \wb x_s$ for $\epsilon \lesssim 0.085$.

Near the merging point, the coupling constants are expected to run slowly and enter a walking regime \cite{Kaplan:2009kr}. It was pointed out in \cite{Gorbenko:2018ncu} that a proper invariant description of the walking regime is provided by the CFT data of so-called complex CFTs, namely the CFTs that appear for imaginary values of the couplings. In our model the fixed points $p_1$ and $p_2$ become complex when $x_s > \wb x_s$ and define two complex CFTs $\cC$ and $\wb\cC$. Theory $\cC$ contains the double-trace operator $\cO^2$ with $\Delta_{\cO^2} = 4 + i |\zeta_3^\star|$, where $\zeta_3^\star = 8\lambda^\star A_+$ is purely imaginary for $x_s > \wb x_s$.
The ``conjugate'' operator with $\Delta_{\cO^2} = 4 - i |\zeta_3^\star|$ is instead in the CFT $\wb\cC$.
We notice that there exists an (unphysical) RG flow for imaginary RG time that connects $\cC$ to $\wb\cC$. This is induced by a deformation $\delta\!f\, \cO^2$ of $\cC$. At large~$N$, $\beta_{\delta\!f} = i |\zeta_3^\star| \, \delta\!f + \alpha \, \delta\!f^2$ (with $\alpha$ a real constant) to all orders in conformal perturbation theory \cite{Gubser:2002vv}. The fixed point at $\delta\!f=0$ is $\cC$ while the one at $\delta\!f=-i |\zeta_3^\star|/\alpha$ is $\wb\cC$ \cite{Gorbenko:2018ncu}.

\begin{figure}[t]
\centering
\includegraphics[height=0.5\columnwidth]{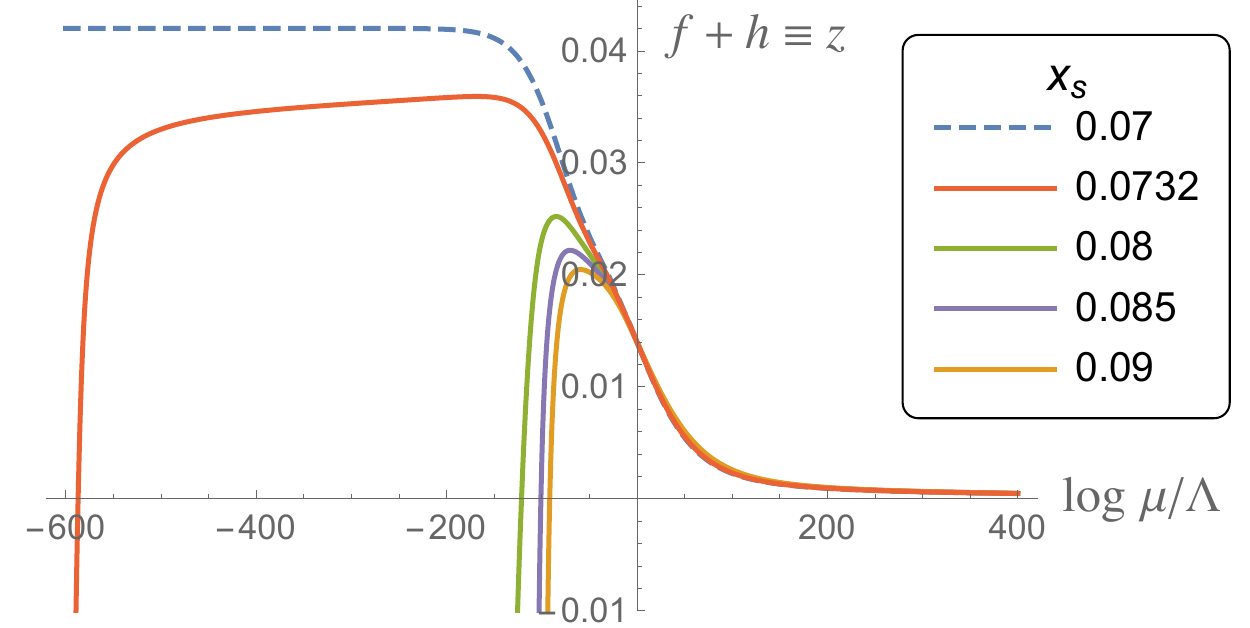}
\caption{RG flow of $f+h$ (controlling stability of the scalar potential) for values of $x_s>\wb x_s$ (the dashed line is for \mbox{$x_s<\wb x_s$}). The theory undergoes spontaneous symmetry breaking via Coleman-Weinberg mechanism around the scale $\mu = \Lambda_\mathrm{IR}$ where $f+h$ vanishes. The initial conditions at $\mu=\Lambda$ are $\lambda= 15 \cdot 10^{-3}$, $h = f = 7 \cdot 10^{-3}$, while $\epsilon=0.04$.
\label{fig:walking}}
\end{figure}

The walking regime is controlled by $|\zeta_3^\star|$ and the RG flow of the double-trace coupling $f$. When $|\zeta_3^\star| \ll 1$, in the walking regime $\lambda \approx \lambda^\star$ and $h\approx h_+^\star$. By redefining $f \to f_0/4 - \lambda^\star B$, the $\beta$-function reads
$\beta_{f_0} \approx f_0^2 + |\zeta_3^\star|^2/4$, with solution $f_0(\mu) \approx |\zeta_3^\star|/2 \tan \bigl( |\zeta_3^\star| /2 \log \mu/\Lambda \bigr)$, $\Lambda$ being an integration constant. The ratio of scales where the walking regime occurs is approximately given by $\exp \bigl( 2\pi/|\zeta_3^\star| \bigr)$
\footnote{Such scaling is often denoted Miransky scaling \cite{Miransky:1984ef}.}.
We illustrate this behavior in Fig.~\ref{fig:walking}, where we plot the RG flow of $f+h$ for various values of $x_s> \wb x_s$, when
there is no IR-stable fixed point and radiative symmetry breaking occurs in the IR through dimensional transmutation. The red line corresponds to the walking regime for $x_s-\wb x_s \ll 1$ (while the dashed blue line is an IR-stable flow for $x_s < \wb x_s$).

When the IR-stable fixed point moves in the complex plane, the RG flow continues until, at some energy scale, all $N_s$ scalars condense with color-flavor locking pattern:
\be 
\label{VEV}
\langle \phi_c^a \rangle = v \, \delta_c^a \;.
\ee
This pattern follows from the fact that the single-trace coupling $h$ is positive when the coupling $z \equiv f+h$ becomes negative \cite{Gildener:1975cj}.
We can determine $v$ by computing the effective potential. For our purposes, it will be enough to approximate it by means of the tree-level RG improved potential, evaluated at the scale $\Lambda_\mathrm{IR}$ where $z = 0$ (see Fig.~\ref{fig:walking}) and no large logs appear. 
Around the vacuum expectation value (VEV) (\ref{VEV}) the double- and single-trace couplings combine into $z$. Around $z\approx 0$ we have
\be
\frac{V_\mathrm{eff}(v)}{8\pi^2} \approx x_s \biggl( 8 x_s h_+^{\star 2} + \frac 34 (1+x_s) \lambda^{\star2} \biggr) \, v^4 \log \frac{v^2}{\Lambda_\mathrm{IR}^2} \;.
\label{Veff}
\ee
This potential has a minimum at $v/\Lambda_\mathrm{IR} = e^{-\frac14}$, where $V_\mathrm{eff}(v)<0$. By adding a tiny UV mass term for the scalars, $m^2 \sim \lambda^{\star2} \Lambda_\mathrm{IR}^2/N_c$, we explicitly see that the phase transition is (weakly) of first order,  which is the typical case occurring in radiative symmetry breaking induced by dimensional transmutation.

The scalar VEV (\ref{VEV}) breaks the gauge group to $SU(N_c-N_s)$.
Scalars acquire a tree-level squared mass of order $h v^2/N_c$, gluons of order $\lambda v^2/N_c$, while the scalar parametrizing the radial fluctuation of $v$ gets a one-loop squared mass schematically of order $(h^2 + \lambda^2) v^2/N_c$ \cite{Gildener:1976ih}. The latter is proportional to the $\beta$-function of $z$ at $z=0$,
which is not parametrically small in the parameter $x_s - \wb x_s$.
In the walking regime, the scale $\Lambda_\mathrm{IR}$ can be made arbitrarily low. However, the gluon-to-scalar mass ratio remains constant.
It has been conjectured, in the context of technicolor models, that a scalar particle---denoted technidilaton---is parametrically lighter
than the other resonances, and corresponds to the pseudo-Nambu-Goldstone boson of the spontaneously broken conformal symmetry \cite{Yamawaki:1985zg}.
In our weakly coupled description of walking, the radial fluctuation is the lightest excitation, but it cannot be made parametrically lighter than the others when $x_s \rightarrow \wb x_s$. This is in agreement with the absence of spontaneous breaking of conformal symmetry, as conformality is broken explicitly.

Let us discuss the effect of higher order corrections and finite $N$ models.
Higher loop corrections to (\ref{beta functions}) correct the location of the fixed points to $g^\star = g^\star_1+ \sum_{l=2}  \delta_l g^\star$, where $g^\star_1=O(\epsilon)$ (for $\epsilon$ sufficiently small) is any of the values in (\ref{lambda star})--(\ref{f star}) while $\delta_l g^\star = O(\epsilon^l)$ are the corrections expected at loop level $l$ in the scalar and $l+1$ in the gauge $\beta$-functions
\footnote{Here $g$ stands for any of the couplings $\lambda$, $h$ and $f$ considered before. Higher loop corrections can also induce new fixed points of the $\beta$-functions. These, however, are not under perturbative control. See \cite{Antipin:2012kc} for a description of walking dynamics based on such fixed points.}.
If we take $N_c \gtrsim O\bigl( \epsilon^{-\frac{l-1}2} \bigr)$ with $l\geq 2$, we can neglect $O(1/N_c^2)$ corrections to the $\beta$-functions up to loop level $l$. So by appropriately choosing the scaling of $\epsilon$, our qualitative analysis is not expected to change up to arbitrary high orders, provided $N_c$ is large enough.

Since the interesting region at large $N_c$ occurs for $N_s \ll N_c$, at finite $N_c$ we consider small $N_s$.  We choose $N_f(N_c)$ such that the theory is the most weakly coupled one and perturbative computations are expected to make sense. The qualitative analysis is the same as in the large $N$ case.
For each value of $N_s$, there is a maximal integer value $N_c^\star$ at and below which the IR-stable fixed point has merged to another one:
conformality is lost for $N_c \leq N_c^\star$. For $N_s=2,3,4$, we find $N_c^\star = 25,39,53$, corresponding to $\wb x_s \simeq 0.08,\, 0.077,\, 0.075$. Note how the large $N$ value $\wb x_s \simeq 0.073$ is quickly approached. For $N_c > N_c^\star$, there are UV-free RG flows  leading to the stable IR fixed point.
For  $N_s=1$ the single- and double-trace quartic couplings combine in the single coupling $z = N_c \bigl( \tilde f + \tilde h \bigr) / 16\pi^2$.
Independently of $N_f$, for $N_c\geq 3$ we always find two real fixed points $z_\pm^\star$ and UV-free RG flows,
while for $N_c=2$ the $z_\pm^\star$ are complex and the theory is not UV free.

Summarizing, we have shown explicit, weakly coupled, and UV complete 4D gauge theories where a conformal window ends by merging of two fixed points and a walking regime appears, after which dynamical symmetry breaking occurs. 
Models of this kind could also have phenomenological applications.
In particular, UV complete gauge theories with walking regime, near the edge of their conformal window, are interesting candidates for possible composite Higgs models.
In the context of natural theories, fundamental scalars should be avoided, and the dynamics should occur at strong coupling, as in QCD. 
Although we have shown that a dilaton-like scalar particle is not parametrically lighter than the other excitations, yet we generally expect it to be the lightest resonance
and as such could be a clear collider signature for models of this kind.

\begin{acknowledgements}
We thank Hrachya Khachatryan for collaboration at the early stages of this project. We also thank Lorenzo Di Pietro and Slava Rychkov for discussions.
F.B. is supported in part by the MIUR-SIR Grant No. RBSI1471GJ.
M.S. acknowledges support from the Simons Collaboration on the Nonperturbative Bootstrap.
\end{acknowledgements}

\bibliography{CFTs_PRL}

\begin{thebibliography}{32}%
\makeatletter
\providecommand \@ifxundefined [1]{%
 \@ifx{#1\undefined}
}%
\providecommand \@ifnum [1]{%
 \ifnum #1\expandafter \@firstoftwo
 \else \expandafter \@secondoftwo
 \fi
}%
\providecommand \@ifx [1]{%
 \ifx #1\expandafter \@firstoftwo
 \else \expandafter \@secondoftwo
 \fi
}%
\providecommand \natexlab [1]{#1}%
\providecommand \enquote  [1]{``#1''}%
\providecommand \bibnamefont  [1]{#1}%
\providecommand \bibfnamefont [1]{#1}%
\providecommand \citenamefont [1]{#1}%
\providecommand \href@noop [0]{\@secondoftwo}%
\providecommand \href [0]{\begingroup \@sanitize@url \@href}%
\providecommand \@href[1]{\@@startlink{#1}\@@href}%
\providecommand \@@href[1]{\endgroup#1\@@endlink}%
\providecommand \@sanitize@url [0]{\catcode `\\12\catcode `\$12\catcode
  `\&12\catcode `\#12\catcode `\^12\catcode `\_12\catcode `\%12\relax}%
\providecommand \@@startlink[1]{}%
\providecommand \@@endlink[0]{}%
\providecommand \url  [0]{\begingroup\@sanitize@url \@url }%
\providecommand \@url [1]{\endgroup\@href {#1}{\urlprefix }}%
\providecommand \urlprefix  [0]{URL }%
\providecommand \Eprint [0]{\href }%
\providecommand \doibase [0]{http://dx.doi.org/}%
\providecommand \selectlanguage [0]{\@gobble}%
\providecommand \bibinfo  [0]{\@secondoftwo}%
\providecommand \bibfield  [0]{\@secondoftwo}%
\providecommand \translation [1]{[#1]}%
\providecommand \BibitemOpen [0]{}%
\providecommand \bibitemStop [0]{}%
\providecommand \bibitemNoStop [0]{.\EOS\space}%
\providecommand \EOS [0]{\spacefactor3000\relax}%
\providecommand \BibitemShut  [1]{\csname bibitem#1\endcsname}%
\let\auto@bib@innerbib\@empty
\bibitem [{\citenamefont {Kaplan}\ \emph {et~al.}(2009)\citenamefont {Kaplan},
  \citenamefont {Lee}, \citenamefont {Son},\ and\ \citenamefont
  {Stephanov}}]{Kaplan:2009kr}%
  \BibitemOpen
  \bibfield  {author} {\bibinfo {author} {\bibfnamefont {D.~B.}\ \bibnamefont
  {Kaplan}}, \bibinfo {author} {\bibfnamefont {J.-W.}\ \bibnamefont {Lee}},
  \bibinfo {author} {\bibfnamefont {D.~T.}\ \bibnamefont {Son}}, \ and\
  \bibinfo {author} {\bibfnamefont {M.~A.}\ \bibnamefont {Stephanov}},\ }\href
  {\doibase 10.1103/PhysRevD.80.125005} {\bibfield  {journal} {\bibinfo
  {journal} {Phys. Rev.}\ }\textbf {\bibinfo {volume} {D80}},\ \bibinfo {pages}
  {125005} (\bibinfo {year} {2009})}\BibitemShut {NoStop}%
\bibitem [{\citenamefont {Gies}\ and\ \citenamefont
  {Jaeckel}(2006)}]{Gies:2005as}%
  \BibitemOpen
  \bibfield  {author} {\bibinfo {author} {\bibfnamefont {H.}~\bibnamefont
  {Gies}}\ and\ \bibinfo {author} {\bibfnamefont {J.}~\bibnamefont {Jaeckel}},\
  }\href {\doibase 10.1140/epjc/s2006-02475-0} {\bibfield  {journal} {\bibinfo
  {journal} {Eur. Phys. J.}\ }\textbf {\bibinfo {volume} {C46}},\ \bibinfo
  {pages} {433} (\bibinfo {year} {2006})}\BibitemShut {NoStop}%
\bibitem [{\citenamefont {Holdom}(1981)}]{Holdom:1981rm}%
  \BibitemOpen
  \bibfield  {author} {\bibinfo {author} {\bibfnamefont {B.}~\bibnamefont
  {Holdom}},\ }\href {\doibase 10.1103/PhysRevD.24.1441} {\bibfield  {journal}
  {\bibinfo  {journal} {Phys. Rev.}\ }\textbf {\bibinfo {volume} {D24}},\
  \bibinfo {pages} {1441} (\bibinfo {year} {1981})}\BibitemShut {NoStop}%
\bibitem [{\citenamefont {Yamawaki}\ \emph {et~al.}(1986)\citenamefont
  {Yamawaki}, \citenamefont {Bando},\ and\ \citenamefont
  {Matumoto}}]{Yamawaki:1985zg}%
  \BibitemOpen
  \bibfield  {author} {\bibinfo {author} {\bibfnamefont {K.}~\bibnamefont
  {Yamawaki}}, \bibinfo {author} {\bibfnamefont {M.}~\bibnamefont {Bando}}, \
  and\ \bibinfo {author} {\bibfnamefont {K.}~\bibnamefont {Matumoto}},\ }\href
  {\doibase 10.1103/PhysRevLett.56.1335} {\bibfield  {journal} {\bibinfo
  {journal} {Phys. Rev. Lett.}\ }\textbf {\bibinfo {volume} {56}},\ \bibinfo
  {pages} {1335} (\bibinfo {year} {1986})}\BibitemShut {NoStop}%
\bibitem [{\citenamefont {Appelquist}\ \emph {et~al.}(1986)\citenamefont
  {Appelquist}, \citenamefont {Karabali},\ and\ \citenamefont
  {Wijewardhana}}]{Appelquist:1986an}%
  \BibitemOpen
  \bibfield  {author} {\bibinfo {author} {\bibfnamefont {T.~W.}\ \bibnamefont
  {Appelquist}}, \bibinfo {author} {\bibfnamefont {D.}~\bibnamefont
  {Karabali}}, \ and\ \bibinfo {author} {\bibfnamefont {L.~C.~R.}\ \bibnamefont
  {Wijewardhana}},\ }\href {\doibase 10.1103/PhysRevLett.57.957} {\bibfield
  {journal} {\bibinfo  {journal} {Phys. Rev. Lett.}\ }\textbf {\bibinfo
  {volume} {57}},\ \bibinfo {pages} {957} (\bibinfo {year} {1986})}\BibitemShut
  {NoStop}%
\bibitem [{\citenamefont {Appelquist}\ \emph {et~al.}(1996)\citenamefont
  {Appelquist}, \citenamefont {Terning},\ and\ \citenamefont
  {Wijewardhana}}]{Appelquist:1996dq}%
  \BibitemOpen
  \bibfield  {author} {\bibinfo {author} {\bibfnamefont {T.}~\bibnamefont
  {Appelquist}}, \bibinfo {author} {\bibfnamefont {J.}~\bibnamefont {Terning}},
  \ and\ \bibinfo {author} {\bibfnamefont {L.~C.~R.}\ \bibnamefont
  {Wijewardhana}},\ }\href {\doibase 10.1103/PhysRevLett.77.1214} {\bibfield
  {journal} {\bibinfo  {journal} {Phys. Rev. Lett.}\ }\textbf {\bibinfo
  {volume} {77}},\ \bibinfo {pages} {1214} (\bibinfo {year}
  {1996})}\BibitemShut {NoStop}%
\bibitem [{Note1()}]{Note1}%
  \BibitemOpen
  \bibinfo {note} {Walking dynamics has been mostly discussed within
  technicolor models (see \cite {Hill:2002ap} for a review and further
  references), but it can play an important role also in its modern
  reincarnation in terms of UV-complete composite Higgs models, as in \cite
  {Ferretti:2013kya}.}\BibitemShut {Stop}%
\bibitem [{\citenamefont {Gorbenko}\ \emph
  {et~al.}(2018{\natexlab{a}})\citenamefont {Gorbenko}, \citenamefont
  {Rychkov},\ and\ \citenamefont {Zan}}]{Gorbenko:2018ncu}%
  \BibitemOpen
  \bibfield  {author} {\bibinfo {author} {\bibfnamefont {V.}~\bibnamefont
  {Gorbenko}}, \bibinfo {author} {\bibfnamefont {S.}~\bibnamefont {Rychkov}}, \
  and\ \bibinfo {author} {\bibfnamefont {B.}~\bibnamefont {Zan}},\ }\href
  {\doibase 10.1007/JHEP10(2018)108} {\bibfield  {journal} {\bibinfo  {journal}
  {JHEP}\ }\textbf {\bibinfo {volume} {10}},\ \bibinfo {pages} {108} (\bibinfo
  {year} {2018}{\natexlab{a}})}\BibitemShut {NoStop}%
\bibitem [{\citenamefont {Gorbenko}\ \emph
  {et~al.}(2018{\natexlab{b}})\citenamefont {Gorbenko}, \citenamefont
  {Rychkov},\ and\ \citenamefont {Zan}}]{Gorbenko:2018dtm}%
  \BibitemOpen
  \bibfield  {author} {\bibinfo {author} {\bibfnamefont {V.}~\bibnamefont
  {Gorbenko}}, \bibinfo {author} {\bibfnamefont {S.}~\bibnamefont {Rychkov}}, \
  and\ \bibinfo {author} {\bibfnamefont {B.}~\bibnamefont {Zan}},\ }\href
  {\doibase 10.21468/SciPostPhys.5.5.050} {\bibfield  {journal} {\bibinfo
  {journal} {SciPost Phys.}\ }\textbf {\bibinfo {volume} {5}},\ \bibinfo
  {pages} {050} (\bibinfo {year} {2018}{\natexlab{b}})}\BibitemShut {NoStop}%
\bibitem [{\citenamefont {Appelquist}\ \emph {et~al.}(1988)\citenamefont
  {Appelquist}, \citenamefont {Nash},\ and\ \citenamefont
  {Wijewardhana}}]{Appelquist:1988sr}%
  \BibitemOpen
  \bibfield  {author} {\bibinfo {author} {\bibfnamefont {T.}~\bibnamefont
  {Appelquist}}, \bibinfo {author} {\bibfnamefont {D.}~\bibnamefont {Nash}}, \
  and\ \bibinfo {author} {\bibfnamefont {L.~C.~R.}\ \bibnamefont
  {Wijewardhana}},\ }\href {\doibase 10.1103/PhysRevLett.60.2575} {\bibfield
  {journal} {\bibinfo  {journal} {Phys. Rev. Lett.}\ }\textbf {\bibinfo
  {volume} {60}},\ \bibinfo {pages} {2575} (\bibinfo {year}
  {1988})}\BibitemShut {NoStop}%
\bibitem [{\citenamefont {Appelquist}\ and\ \citenamefont
  {Nash}(1990)}]{Appelquist:1989tc}%
  \BibitemOpen
  \bibfield  {author} {\bibinfo {author} {\bibfnamefont {T.}~\bibnamefont
  {Appelquist}}\ and\ \bibinfo {author} {\bibfnamefont {D.}~\bibnamefont
  {Nash}},\ }\href {\doibase 10.1103/PhysRevLett.64.721} {\bibfield  {journal}
  {\bibinfo  {journal} {Phys. Rev. Lett.}\ }\textbf {\bibinfo {volume} {64}},\
  \bibinfo {pages} {721} (\bibinfo {year} {1990})}\BibitemShut {NoStop}%
\bibitem [{\citenamefont {Nahum}\ \emph {et~al.}(2015)\citenamefont {Nahum},
  \citenamefont {Chalker}, \citenamefont {Serna}, \citenamefont {Ortu\~{n}o},\
  and\ \citenamefont {Somoza}}]{Nahum:2015jya}%
  \BibitemOpen
  \bibfield  {author} {\bibinfo {author} {\bibfnamefont {A.}~\bibnamefont
  {Nahum}}, \bibinfo {author} {\bibfnamefont {J.~T.}\ \bibnamefont {Chalker}},
  \bibinfo {author} {\bibfnamefont {P.}~\bibnamefont {Serna}}, \bibinfo
  {author} {\bibfnamefont {M.}~\bibnamefont {Ortu\~{n}o}}, \ and\ \bibinfo
  {author} {\bibfnamefont {A.~M.}\ \bibnamefont {Somoza}},\ }\href {\doibase
  10.1103/PhysRevX.5.041048} {\bibfield  {journal} {\bibinfo  {journal} {Phys.
  Rev.}\ }\textbf {\bibinfo {volume} {X5}},\ \bibinfo {pages} {041048}
  (\bibinfo {year} {2015})}\BibitemShut {NoStop}%
\bibitem [{Note2()}]{Note2}%
  \BibitemOpen
  \bibinfo {note} {Stability means that the scalar potential is bounded below,
  to give a consistent and unitary quantum theory.}\BibitemShut {Stop}%
\bibitem [{\citenamefont {Gubser}\ and\ \citenamefont
  {Klebanov}(2003)}]{Gubser:2002vv}%
  \BibitemOpen
  \bibfield  {author} {\bibinfo {author} {\bibfnamefont {S.~S.}\ \bibnamefont
  {Gubser}}\ and\ \bibinfo {author} {\bibfnamefont {I.~R.}\ \bibnamefont
  {Klebanov}},\ }\href {\doibase 10.1016/S0550-3213(03)00056-7} {\bibfield
  {journal} {\bibinfo  {journal} {Nucl. Phys.}\ }\textbf {\bibinfo {volume}
  {B656}},\ \bibinfo {pages} {23} (\bibinfo {year} {2003})}\BibitemShut
  {NoStop}%
\bibitem [{\citenamefont {Witten}(2001)}]{Witten:2001ua}%
  \BibitemOpen
  \bibfield  {author} {\bibinfo {author} {\bibfnamefont {E.}~\bibnamefont
  {Witten}},\ }\href@noop {} {\bibfield  {journal} {\bibinfo  {journal}
  {arXiv:hep-th/0112258}\ } (\bibinfo {year} {2001})}\BibitemShut {NoStop}%
\bibitem [{\citenamefont {Dymarsky}\ \emph {et~al.}(2005)\citenamefont
  {Dymarsky}, \citenamefont {Klebanov},\ and\ \citenamefont
  {Roiban}}]{Dymarsky:2005uh}%
  \BibitemOpen
  \bibfield  {author} {\bibinfo {author} {\bibfnamefont {A.}~\bibnamefont
  {Dymarsky}}, \bibinfo {author} {\bibfnamefont {I.~R.}\ \bibnamefont
  {Klebanov}}, \ and\ \bibinfo {author} {\bibfnamefont {R.}~\bibnamefont
  {Roiban}},\ }\href {\doibase 10.1088/1126-6708/2005/08/011} {\bibfield
  {journal} {\bibinfo  {journal} {JHEP}\ }\textbf {\bibinfo {volume} {08}},\
  \bibinfo {pages} {011} (\bibinfo {year} {2005})}\BibitemShut {NoStop}%
\bibitem [{\citenamefont {Pomoni}\ and\ \citenamefont
  {Rastelli}(2009)}]{Pomoni:2008de}%
  \BibitemOpen
  \bibfield  {author} {\bibinfo {author} {\bibfnamefont {E.}~\bibnamefont
  {Pomoni}}\ and\ \bibinfo {author} {\bibfnamefont {L.}~\bibnamefont
  {Rastelli}},\ }\href {\doibase 10.1088/1126-6708/2009/04/020} {\bibfield
  {journal} {\bibinfo  {journal} {JHEP}\ }\textbf {\bibinfo {volume} {04}},\
  \bibinfo {pages} {020} (\bibinfo {year} {2009})}\BibitemShut {NoStop}%
\bibitem [{Note3()}]{Note3}%
  \BibitemOpen
  \bibinfo {note} {The possibility of a walking behavior was discussed in \cite
  {Dymarsky:2005uh} in the context of orbifold models of ${\protect \cal N}=4$
  supersymmetric Yang-Mills theory, where $b(\lambda )$ in (\ref {betaflambda})
  vanishes identically. However, no real fixed points were found, and such
  theories are generally not UV complete.}\BibitemShut {Stop}%
\bibitem [{Note4()}]{Note4}%
  \BibitemOpen
  \bibinfo {note} {A numerical study of this theory at finite $N$, including
  its RG flow, regions of conformality and UV freedom, already appeared in
  \cite {Hansen:2017pwe}. However, an analytic large $N$ study, description of
  fixed-point merging and walking behaviour was not provided
  there.}\BibitemShut {Stop}%
\bibitem [{\citenamefont {Caswell}(1974)}]{Caswell:1974gg}%
  \BibitemOpen
  \bibfield  {author} {\bibinfo {author} {\bibfnamefont {W.~E.}\ \bibnamefont
  {Caswell}},\ }\href {\doibase 10.1103/PhysRevLett.33.244} {\bibfield
  {journal} {\bibinfo  {journal} {Phys. Rev. Lett.}\ }\textbf {\bibinfo
  {volume} {33}},\ \bibinfo {pages} {244} (\bibinfo {year} {1974})}\BibitemShut
  {NoStop}%
\bibitem [{\citenamefont {Banks}\ and\ \citenamefont
  {Zaks}(1982)}]{Banks:1981nn}%
  \BibitemOpen
  \bibfield  {author} {\bibinfo {author} {\bibfnamefont {T.}~\bibnamefont
  {Banks}}\ and\ \bibinfo {author} {\bibfnamefont {A.}~\bibnamefont {Zaks}},\
  }\href {\doibase 10.1016/0550-3213(82)90035-9} {\bibfield  {journal}
  {\bibinfo  {journal} {Nucl. Phys.}\ }\textbf {\bibinfo {volume} {B196}},\
  \bibinfo {pages} {189} (\bibinfo {year} {1982})}\BibitemShut {NoStop}%
\bibitem [{\citenamefont {Machacek}\ and\ \citenamefont
  {Vaughn}(1983)}]{Machacek:1983tz}%
  \BibitemOpen
  \bibfield  {author} {\bibinfo {author} {\bibfnamefont {M.~E.}\ \bibnamefont
  {Machacek}}\ and\ \bibinfo {author} {\bibfnamefont {M.~T.}\ \bibnamefont
  {Vaughn}},\ }\href {\doibase 10.1016/0550-3213(83)90610-7} {\bibfield
  {journal} {\bibinfo  {journal} {Nucl. Phys.}\ }\textbf {\bibinfo {volume}
  {B222}},\ \bibinfo {pages} {83} (\bibinfo {year} {1983})}\BibitemShut
  {NoStop}%
\bibitem [{\citenamefont {Machacek}\ and\ \citenamefont
  {Vaughn}(1985)}]{Machacek:1984zw}%
  \BibitemOpen
  \bibfield  {author} {\bibinfo {author} {\bibfnamefont {M.~E.}\ \bibnamefont
  {Machacek}}\ and\ \bibinfo {author} {\bibfnamefont {M.~T.}\ \bibnamefont
  {Vaughn}},\ }\href {\doibase 10.1016/0550-3213(85)90040-9} {\bibfield
  {journal} {\bibinfo  {journal} {Nucl. Phys.}\ }\textbf {\bibinfo {volume}
  {B249}},\ \bibinfo {pages} {70} (\bibinfo {year} {1985})}\BibitemShut
  {NoStop}%
\bibitem [{\citenamefont {Hansen}\ \emph {et~al.}(2018)\citenamefont {Hansen},
  \citenamefont {Janowski}, \citenamefont {Lang\ae{}ble}, \citenamefont {Mann},
  \citenamefont {Sannino}, \citenamefont {Steele},\ and\ \citenamefont
  {Wang}}]{Hansen:2017pwe}%
  \BibitemOpen
  \bibfield  {author} {\bibinfo {author} {\bibfnamefont {F.~F.}\ \bibnamefont
  {Hansen}}, \bibinfo {author} {\bibfnamefont {T.}~\bibnamefont {Janowski}},
  \bibinfo {author} {\bibfnamefont {K.}~\bibnamefont {Lang\ae{}ble}}, \bibinfo
  {author} {\bibfnamefont {R.~B.}\ \bibnamefont {Mann}}, \bibinfo {author}
  {\bibfnamefont {F.}~\bibnamefont {Sannino}}, \bibinfo {author} {\bibfnamefont
  {T.~G.}\ \bibnamefont {Steele}}, \ and\ \bibinfo {author} {\bibfnamefont
  {Z.-W.}\ \bibnamefont {Wang}},\ }\href {\doibase 10.1103/PhysRevD.97.065014}
  {\bibfield  {journal} {\bibinfo  {journal} {Phys. Rev.}\ }\textbf {\bibinfo
  {volume} {D97}},\ \bibinfo {pages} {065014} (\bibinfo {year}
  {2018})}\BibitemShut {NoStop}%
\bibitem [{Note5()}]{Note5}%
  \BibitemOpen
  \bibinfo {note} {Such scaling is often denoted Miransky scaling \cite
  {Miransky:1984ef}.}\BibitemShut {Stop}%
\bibitem [{\citenamefont {Gildener}(1976)}]{Gildener:1975cj}%
  \BibitemOpen
  \bibfield  {author} {\bibinfo {author} {\bibfnamefont {E.}~\bibnamefont
  {Gildener}},\ }\href {\doibase 10.1103/PhysRevD.13.1025} {\bibfield
  {journal} {\bibinfo  {journal} {Phys. Rev.}\ }\textbf {\bibinfo {volume}
  {D13}},\ \bibinfo {pages} {1025} (\bibinfo {year} {1976})}\BibitemShut
  {NoStop}%
\bibitem [{\citenamefont {Gildener}\ and\ \citenamefont
  {Weinberg}(1976)}]{Gildener:1976ih}%
  \BibitemOpen
  \bibfield  {author} {\bibinfo {author} {\bibfnamefont {E.}~\bibnamefont
  {Gildener}}\ and\ \bibinfo {author} {\bibfnamefont {S.}~\bibnamefont
  {Weinberg}},\ }\href {\doibase 10.1103/PhysRevD.13.3333} {\bibfield
  {journal} {\bibinfo  {journal} {Phys. Rev. D}\ }\textbf {\bibinfo {volume}
  {13}},\ \bibinfo {pages} {3333} (\bibinfo {year} {1976})}\BibitemShut
  {NoStop}%
\bibitem [{Note6()}]{Note6}%
  \BibitemOpen
  \bibinfo {note} {Here $g$ stands for any of the couplings $\lambda $, $h$ and
  $f$ considered before. Higher loop corrections can also induce new fixed
  points of the $\beta $-functions. These, however, are not under perturbative
  control. See \cite {Antipin:2012kc} for a description of walking dynamics
  based on such fixed points.}\BibitemShut {Stop}%
\bibitem [{\citenamefont {Hill}\ and\ \citenamefont
  {Simmons}(2003)}]{Hill:2002ap}%
  \BibitemOpen
  \bibfield  {author} {\bibinfo {author} {\bibfnamefont {C.~T.}\ \bibnamefont
  {Hill}}\ and\ \bibinfo {author} {\bibfnamefont {E.~H.}\ \bibnamefont
  {Simmons}},\ }\href {\doibase 10.1016/S0370-1573(03)00140-6} {\bibfield
  {journal} {\bibinfo  {journal} {Phys. Rept.}\ }\textbf {\bibinfo {volume}
  {381}},\ \bibinfo {pages} {235} (\bibinfo {year} {2003})}\BibitemShut
  {NoStop}%
\bibitem [{\citenamefont {Ferretti}\ and\ \citenamefont
  {Karateev}(2014)}]{Ferretti:2013kya}%
  \BibitemOpen
  \bibfield  {author} {\bibinfo {author} {\bibfnamefont {G.}~\bibnamefont
  {Ferretti}}\ and\ \bibinfo {author} {\bibfnamefont {D.}~\bibnamefont
  {Karateev}},\ }\href {\doibase 10.1007/JHEP03(2014)077} {\bibfield  {journal}
  {\bibinfo  {journal} {JHEP}\ }\textbf {\bibinfo {volume} {03}},\ \bibinfo
  {pages} {077} (\bibinfo {year} {2014})}\BibitemShut {NoStop}%
\bibitem [{\citenamefont {Miransky}(1985)}]{Miransky:1984ef}%
  \BibitemOpen
  \bibfield  {author} {\bibinfo {author} {\bibfnamefont {V.~A.}\ \bibnamefont
  {Miransky}},\ }\href {\doibase 10.1007/BF02724229} {\bibfield  {journal}
  {\bibinfo  {journal} {Nuovo Cim.}\ }\textbf {\bibinfo {volume} {A90}},\
  \bibinfo {pages} {149} (\bibinfo {year} {1985})}\BibitemShut {NoStop}%
\bibitem [{\citenamefont {Antipin}\ \emph {et~al.}(2012)\citenamefont
  {Antipin}, \citenamefont {Di~Chiara}, \citenamefont {Mojaza}, \citenamefont
  {M\o{}lgaard},\ and\ \citenamefont {Sannino}}]{Antipin:2012kc}%
  \BibitemOpen
  \bibfield  {author} {\bibinfo {author} {\bibfnamefont {O.}~\bibnamefont
  {Antipin}}, \bibinfo {author} {\bibfnamefont {S.}~\bibnamefont {Di~Chiara}},
  \bibinfo {author} {\bibfnamefont {M.}~\bibnamefont {Mojaza}}, \bibinfo
  {author} {\bibfnamefont {E.}~\bibnamefont {M\o{}lgaard}}, \ and\ \bibinfo
  {author} {\bibfnamefont {F.}~\bibnamefont {Sannino}},\ }\href {\doibase
  10.1103/PhysRevD.86.085009} {\bibfield  {journal} {\bibinfo  {journal} {Phys.
  Rev.}\ }\textbf {\bibinfo {volume} {D86}},\ \bibinfo {pages} {085009}
  (\bibinfo {year} {2012})}\BibitemShut {NoStop}%
\end{thebibliography}%
\end{document}